\documentclass{article}

\usepackage{graphicx} 
\usepackage{ulem}
\usepackage{bm}

\title{\large Finite Size Polyelectrolyte Bundles at Thermodynamic Equilibrium}

\author{Mehmet Sayar$^{*\dagger}$ and Christian Holm$^{\dagger\ddagger}$ \\
  $*$ Koc University, College of Engineering, Istanbul, Turkey\\
  $\dagger$ Max-Planck-Institut f\"{u}r Polymerforschung, Mainz, Germany\\
  $\ddagger$ FIAS, J.W. Goethe - Universit\"{a}t, Frankfurt, Germany\\
}

\date{\today}

\begin{document}

\maketitle

\begin{abstract}
  We present the results of extensive computer simulations performed
  on solutions of monodisperse charged rod-like polyelectrolytes in the
  presence of trivalent counterions. To overcome energy barriers we
  used a combination of parallel tempering and hybrid Monte Carlo
  techniques. Our results show that for small values of the
  electrostatic interaction the solution mostly consists of dispersed
  single rods. The potential of mean force between the polyelectrolyte
  monomers yields an attractive interaction at short distances. For a
  range of larger values of the Bjerrum length, we find finite size
  polyelectrolyte bundles at thermodynamic equilibrium.  Further
  increase of the Bjerrum length eventually leads to phase separation
  and precipitation. We discuss the origin of the observed
  thermodynamic stability of the finite size aggregates.

\end{abstract}

\section{\label{intro}Introduction:
}
Aggregation phenomena in charged macromolecular systems are well known
and have been the focus of many investigations, see recent
theoretical reviews
\cite{holm01a,grosberg02a,levin02a,boroudjerdi05a}. Here, we are
interested in understanding the aggregation behavior of rod-like
charged polyelectrolytes, such as DNA \cite{bloomfield91a},
F-actin\cite{tang96a,kas96a}, and microtubules\cite{tang96a} that
self-organize into bundles under the influence of only electrostatic
interactions. Probably the most investigated systems are solutions of
DNA in the presence of multivalent counterions
( see Refs. in \cite{bloomfield96a}),
which form a variety of different morphologies such as single or
multiple toroids, or bundle-like aggregates.
The origin of the attraction are short range ionic correlations of 
multivalent counterions, a phenomenon that cannot be explained on the
level of mean-field theories like Poisson-Boltzmann
\cite{holm01a,grosberg02a,levin02a,boroudjerdi05a}. Simulations have
already demonstrated this effect a long time ago \cite{guldbrand86a},
but with the uprise of biological physics, it has regained 
interest in recent years.  Many theoretical and computational studies
have dealt with the origin of the attraction, and we have an
excellent qualitative understanding of it (see Refs. in
\cite{holm01a,grosberg02a,levin02a,boroudjerdi05a}).

In a solution of many macromolecules, where multi-body effects are present, it
has been difficult to obtain accurate enough analytical insight into the
aggregate size distribution.  The aggregate size must be determined by the
competition between the surface tension (due to short range attractions), the
repulsive self-energy of the backbone charges, and the entropic degrees of
freedom of the counterions. These effects are all coupled, and therefore pose
a considerable challenge to analytical approaches. For DNA, it has been argued
\cite{ha98a,levin02a,ha99c,stilck02a} that the observed finite size of the
aggregates is a kinetic effect based on the formation of an energy barrier,
and not a consequence of equilibrium thermodynamic properties of the system.
Experiments performed with viral DNA support this conjecture at least for the
investigated systems \cite{lambert00a}. However, recently Henle and Pincus
\cite{henle05} argued that, depending on the actual parameters of the system,
finite or infinite bundles should be possible in the presence of short-ranged
attractions. Treating the system as consisting of sticky charged rods brings
up the analogy to the splitting of a Rayleigh charged hydrophobic droplet,
where the mean-field model has already been solved by Deserno
\cite{deserno01f}.  He observes that the droplet size is always finite, if the
counterions cannot penetrate into the droplet, but that allowing counterion
penetration leads either to finite or infinite droplet sizes, depending on the
parameters. There have been relatively few simulations on bundle formations,
notable exception being the simulations by Stevens
\cite{stevens99a,stevens01a}, showing the possibility that multivalent ions
alone can lead to bundle formations, and the work of Borukhov et al.
\cite{borukhov02a}, which focuses on two-rod systems bundled via short range
mobile linker interactions. In the present work we investigate the bundle
formation for a system of monodisperse charged semi-flexible polymers in the
presence of trivalent counterions. This system was chosen since it is known to
have short range attractions if the electrostatic interactions are
sufficiently large. We investigate the aggregate size distribution as a
function of interaction strength, and demonstrate clearly the existence of
thermodynamically stable finite size aggregates.

Note that there are also synthetic rod-like polyelectrolytes, such as
poly(para-phenylene) oligomers \cite{bockstaller00a}, that form stable
well defined cylindrical micelles due to hydrophobic interaction of
their side chains\cite{limbach04a,limbach05a}. Even though the dominant cause of
attraction is of non-electrostatic origin, their stability diagram has
some relation to the DNA-like systems under investigation which is
clear if one considers an Ansatz like in Ref. \cite{deserno01f}.

In the first section we will present the model for our computational
approach. In the following results and discussion section we will
present the phase diagram for this system, and explain the origin of
the stability of the finite size aggregates. We end with our
conclusions.

\section{\label{model}Model:
}

In this study a coarse-grained representation of polyelectrolyte (PE)
chains and trivalent counterions is used. All particles interact with
purely repulsive Lennard-Jones interactions (Eqn. \ref{lennard-jones})
with cut-off distance $r_{cut}/\sigma=2^{1/6}$, interaction strength
$\epsilon_{LJ}=k_BT$, and the same diameter $\sigma$. Each PE chain is
composed of 30 negatively charged beads connected with FENE bonds
(Eqn. \ref{fene}) with a spring constant of $k_F=7k_BT$ and cut-off
distance $R_{F}/\sigma=2$.  The PE chains are semi-flexible with a
harmonic bending potential (Eqn. \ref{angle}) of stiffness
$k_{\theta}=100 k_BT$. We are interested in the dilute regime where
the bundle-bundle contacts could be ignored. The PE monomer number
density is fixed for all simulations as $\rho/ \sigma^{-3}=7.5
~10^{-5}~$. There are a total number of 61 PE chains and 10 trivalent
counterions per chain in the cubic simulation box, where periodic
boundary conditions were used. The unscreened coulomb interactions
(Eqn. \ref{coulomb}) are calculated using a version of the P3M
method\cite{deserno98a}. In Eqn. \ref{coulomb}, $\lambda_B$ is the
Bjerrum length and is used to control the strength of the
electrostatic interactions. We used MD simulations with a time step of
$\Delta t = 0.005 \tau$, where $\tau$ is the usual Lennard-Jones time
unit, using the Espresso package\cite{limbach06a}.

\begin{equation}
\label{lennard-jones} U_{LJ}(r)= 4 \epsilon_{LJ} \left(
\left(\frac {\sigma} {r} \right)^{12} - \left(\frac {\sigma} {r}
\right)^{6}\right) \textrm{ for } r < r_{cut}
\end{equation}

\begin{equation}
\label{fene} U_{F}(r)= \frac {1} {2} k_{F} R^2_{F}
\ln\left(1-\left(\frac {r} {R_{F}}\right)^2\right) \textrm{ for } r < R_F
\end{equation}

\begin{equation}
\label{angle} U_{\theta}(r)=  \frac {1} {2} k_{\theta} \theta^2
\end{equation}

\begin{equation}
\label{coulomb} U_{C}(r_{ij})= \lambda_B k_B T \frac {q_i q_j} {r_{ij}}
\end{equation}

The simulation of stiff-chain PEs with trivalent counterions possess a
multitude of difficulties because of the long-range electrostatic interactions
and energy barriers. In our previous study\cite{limbach04a} on bundles of
hydrophobically modified PEs, we have shown that one can obtain a phase
diagram by studying the stability of different size aggregates as a function
of hydrophobic interaction strength and Bjerrum length. In that study, MD
simulations with a Langevin thermostat were used to look at the stability of a
preformed assembly of the PEs.

Simulations of the current PE system without explicit hydrophobic
interactions with the same approach as in Ref. \cite{limbach04a} showed
that, for low values of the Bjerrum length
($\lambda_B/\sigma \le 1.70$), bundles are not stable. The initial bundle
quickly falls apart. At equilibrium one occasionally observes short-lived
dimers and trimers, but bigger aggregates are not observed. On the other hand
for $\lambda_B/\sigma \ge 1.80$ the situation changes drastically. If one
starts from aggregates up to size $N=12$ ($N$ is the number of PEs in the
bundle), a fraction of PEs split up, but a finite size aggregate remains in
solution. The size of the remaining aggregate depends on the size of the
starting bundle. However, as the aggregate size is set to $N \ge 19$ the
bundles remain stable throughout the simulation. Beyond $\lambda_B/\sigma=2.0$
independent of the bundle size, no PEs disintegrate from the start-up bundle.
The reason for this interesting behavior could be explained by two different
scenarios. By looking at this as a nucleation problem one can argue that for
$\lambda_B/\sigma=1.8-2.0$ the critical nucleus size is larger than 19
molecules. Since the bundles we have studied are smaller than the critical
nucleus size the bundles never fall apart completely or grow bigger, but one
observes large fluctuations in the bundle size. We have tested this by
studying an aggregate of size 25 with additional dispersed rods in solution
for $\lambda_B/\sigma>2.0$. After an integration time of $5 ~10^4 \tau$ no
aggregate growth is observed, which leads us to conclude that this is not just
a nucleation phenomenon.

Another explanation for this behavior could be high energy barriers for the
PEs to split up. This led us to believe that the Langevin thermostat is not
efficient enough to overcome such energy barriers, which are highly dependent
on $\lambda_B$ and aggregate size. The parallel tempering method (PT)
\cite{wang1986,frenkel02b} with Bjerrum length as the tempering parameter was
chosen to overcome such energy barriers. We have set up 49 ensembles with
$\lambda_B/ \sigma$ ranging from $1.50-2.19$. For the simulation of the
individual ensembles, we have employed a hybrid Monte Carlo method
(HMC)\cite{Duane1987,Irback1994}. We integrated for 50 $\tau$ for each HMC trial,
which yields a good combination of acceptance ratio and efficient sampling of
the configurational space. For each PT exchange attempt, we have done 9 HMC
moves, which we call one cycle. The acceptance ratio of the HMC moves largely
depends on the aggregate size distribution, and decreases rapidly for large
bundle sizes. In order to test the efficiency of the PT+HMC method for the PE
chains, we have conducted two separate simulations with 19 molecules each. The
starting configurations were chosen as a single bundle for the first
simulation and randomly distributed PEs for the second. After both of the
PT+HMC simulations are equilibrated, an identical aggregate size distribution
is observed. The results presented in the following section are obtained from
PT+HMC simulation with 61 PEs and neutralizing trivalent counterions. All
ensembles are started from a configuration of randomly distributed rods. The
set of ensembles have been equilibrated for up to 3500 cycles, and data is
collected from an additional run of at least 2500 cycles, which took roughly
three cpu years on Intel Xeon processors (2.4 GHz).\\

\section{\label{result}Results and Discussion:
}

\begin{figure}
  \begin{center}
    \includegraphics[width=9cm]{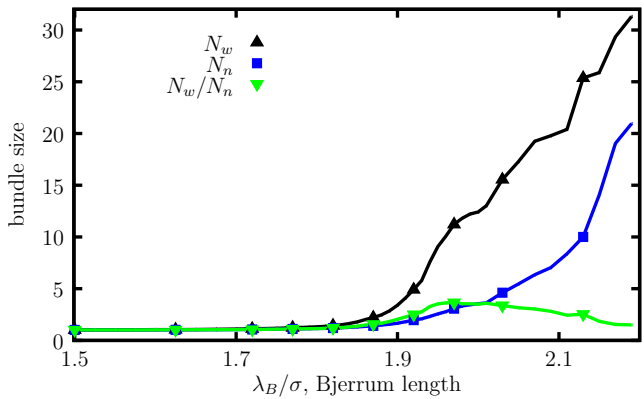}
    \caption{\label{aggsize} The weighted ($N_w$) and number average
      ($N_n$) bundle size as a function of Bjerrum length
      ($\lambda_B$). For $\lambda_B/\sigma \approx 1.80-2.00$ bundles
      of finite size are observed at thermal equilibrium. The $N_w$
      increases with $\lambda_B$ and the polydispersity index has a
      maximum at $\lambda_B/\sigma\approx 1.95$.}
  \end{center}
\end{figure}
We will start our analysis with the aggregate size distribution obtained from
the PT+HMC simulation for the range of Bjerrum length values
($\lambda_B/\sigma=1.50-2.19$). In order to define the aggregation state of
the system, we assumed that any trivalent counterion closer than $3\sigma$ is
condensed onto a PE. If two PEs share one or more condensed counterions they
are assumed to be part of the same bundle. The number average ($N_n=\sum N_i
f_i$) and weighted average ($N_w=\sum N_i^2 f_i / \sum N_i f_i$) aggregate
size (where $f_i$ is the fraction of aggregates of size $N_i$), as well as the
polydispersity index ($N_w/N_n$) is shown in Fig. \ref{aggsize}. In the weak
electrostatic regime ($\lambda_B/\sigma \rightarrow 0$) one observes only
single PEs in solution, and therefore the polydispersity index is exactly one.
For low $\lambda_B$ (up to $\lambda_B/\sigma=1.80$) mostly single
dispersed rods and occasionally short-lived dimers and trimers can be
seen in solution, in agreement with our earlier observations. At such
low values of $\lambda_B$, even though the net charge of the rods is
highly reduced by the condensed counterions, the repulsive
interactions among PEs dominate the system behavior, and the 
polydispersity index is very close to one.
Beyond $\lambda_B/\sigma \approx 1.80$ a gradual increase in the average
aggregate size is observed.  This 
clearly shows that finite size PE bundles exist at thermodynamic equilibrium
for a range of $\lambda_B$ values. In this range the electrostatic
interactions provide both the glue required for aggregation through short
range correlations, and also the repulsive forces, which terminate aggregate
growth. The entropic penalty  still keeps counterions in solution and prevents complete phase separation.  Upon aggregation the polydispersity index rises above one, and shows a peak value of 3.7 around
$\lambda_B/\sigma=1.97$. 
In the high $\lambda_B$ regime ($\lambda_B/\sigma \ge 2.1  $) the correlational attractions dominate,
the entropic terms become negligible, and one expects the formation of
an infinitely large bundle with a dilute solution of PEs.  Therefore
the polydispersity index should go to one again, which it does. Since we have only 61 PEs in our simulation box,
once the average aggregate size reaches above 20, finite size effects
dominate, and one cannot obtain a conclusive answer for the actual average
aggregate size. 

\begin{figure}
  \begin{center}
    \includegraphics[width=9cm]{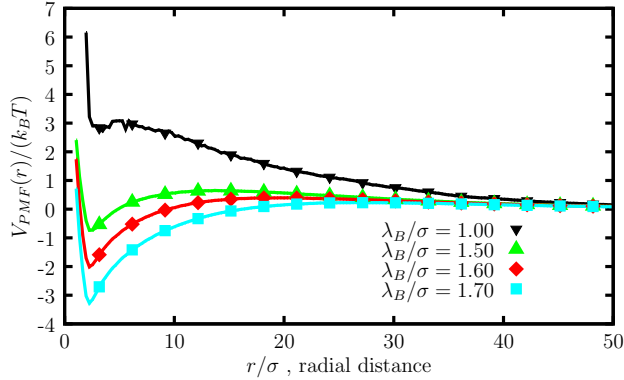}
    \caption{\label{pmf} The potential of mean force among PE
      monomers. The purely repulsive interaction
      ($\lambda_B/\sigma=1.0$) is replaced with a weak attraction
      already for $\lambda_B/\sigma=1.5$. A small energy barrier is
      observed at a finite distance. As $\lambda_B$ is increased
      further the attraction becomes stronger and the energy barrier
      disappears, leading to the onset of aggregation into PE
      bundles.}
  \end{center}
\end{figure}
One can observe the onset of aggregation by looking at the potential of mean
force (PMF) among the centers of mass of the rods. However, the  PMF of two rods
still depends on the relative orientations \cite{ha99c,lee04a}. Instead, we
have chosen to look at the PMF of the PE monomers (Fig. \ref{pmf}), which is
obtained by inverting their radial distribution function.  We have performed
an extra simulation with $\lambda_B/\sigma=1.0$ to demonstrate the purely
repulsive regime. Already for $\lambda_B/\sigma= 1.5$, the effective charge of
the PEs is highly reduced and at short distances a weak attractive interaction
($\approx k_BT$) is observed. As seen in Fig. \ref{aggsize} this weak
attraction is not sufficient to keep even two PEs together. The presence of a
weak energy barrier is also observed for $\lambda_B/\sigma= 1.5$. As one
increases $\lambda_B$ further the energy barrier disappears, and the binding
energy due to counterion correlations is significantly increased, leading to
the formation of stable finite size aggregates.

\begin{figure}
  \begin{center}
    \includegraphics[width=9cm]{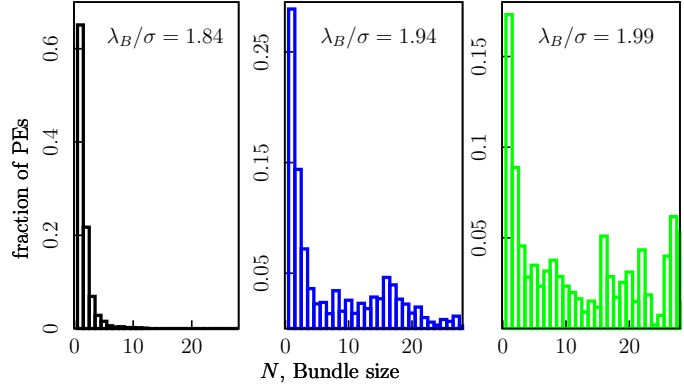}
    \caption{\label{aggdist} The distribution of PEs among different size
      aggregates. Beyond the critical $\lambda_B$ a micellization type size
      distribution is observed, where monomers are in equilibrium with finite
      size PE bundles.}
  \end{center}
\end{figure}
The distribution of PEs among different size aggregates at fixed $\lambda_B$
reveals that this is indeed a micellization phenomenon\cite{huang97} (Fig.
\ref{aggdist}). For $\lambda_B/\sigma=1.84$ more than $60\%$ of the PEs exist
as dispersed single rods, whereas long-lived dimers and trimers exist in
equilibrium with the single PEs. Upon further increase of
($\lambda_B/\sigma=1.94$) a peak for a finite aggregate size $N\approx17$ is
observed. At $\lambda_B/\sigma=1.99$ the distribution broadens, and no single
distinguishable peak remains.
\begin{figure}
  \begin{center}
    \includegraphics[width=9cm]{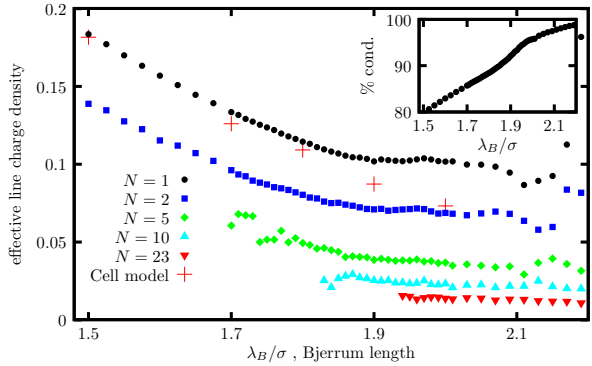}
    \caption{\label{linecharge2} Effective line charge density ($f_e$)
      as a function of $\lambda_B$ for aggregates composed of $N$
      rods. At $\lambda_B \approx 1.8 \sigma$ the average aggregate
      size strongly increases, which leads to a flattening of
      $f_e$. For comparison we show $f_e$ of a single PE in the cell
      model, where multi-body effects are not present. In the insert
      the total fraction of condensed counterions is shown as a
      function of $\lambda_B$.}
  \end{center}
\end{figure}

Next, we will look at the effective line charge density ($f_e$) of different
size bundles. For calculating $f_e$, we assume that a bundle can be
approximated as a rod-like charged object. The net charge of the bundle (PEs
and condensed counterions) is divided by the length of the PEs to obtain
$f_e$.  Note that these bundles have a finite diameter and small aspect ratio.
Furthermore, the PEs are not perfectly aligned, but one observes large
fluctuations along the bundle axis. In Fig. \ref{linecharge2} we plot $f_e$
for aggregates of size 1, 2, 5, 10, and 23 as a function of $\lambda_B$. For
single PEs $f_e$ decreases linearly up to $\lambda_B/\sigma\approx1.8$. Beyond
this value we observe a quick convergence of $f_e$ to a plateau value of
$\approx 0.1$. If we look at $f_e$ obtained for a single PE simulated at the
same density within the cell model, no such convergence is observed. The cell
model simulations match with many-rod PT+HMC simulations only for low
$\lambda_B$. This mismatch demonstrates the importance of many body effects,
such as aggregation of PEs into bundles.
For the many-rod simulation we see that beyond $\lambda_B/\sigma>1.8$
most of the single PEs compensate a high fraction of their intrinsic
charge via condensing counterions. As a result the repulsion among PEs
is greatly diminished and the single PEs with the lowest line charge
density aggregate to form bundles, leaving behind only the single PEs
with higher $f_e$.  Upon aggregation of PEs, the trivalent counterions
gain enthalpy, without diminishing the system entropy further
more. Therefore $f_e$ for single PEs deviates from the cell model. For
aggregates of size of 2, 5, 10, and 23 a similar behavior is
observed. The reduction in $f_e$ continues until merging into even
bigger aggregates becomes feasible, lowering the free energy even more. Note
that large aggregate sizes are only observed beyond a certain
$\lambda_B$ value.
For example, for $N=5$ the high fluctuations in $f_e$ around
$\lambda_B/\sigma\approx 1.7$ stem from the fact that these aggregates
are very rare at such $\lambda_B$ values. The insert in
Fig. \ref{linecharge2} denotes the total fraction of condensed
counterions over all PEs. This fraction increases linearly up to
$\lambda_B/\sigma\approx1.8$. Beyond $\lambda_B/\sigma\approx2.0$ the
charge compensation begins to saturate. Note that at this point more
than $95\%$ of the charge on the PEs are compensated, and only a small
fraction of counterions are left in solution. Therefore the entropic
penalty associated with condensing these remaining free counterions is
extremely high. The deviation from the linear regime nicely coincides
with the onset of aggregation.


If we want to map our results to biomolecules such as DNA in a
trivalent counterion solution we can do so by requiring that the
strong coupling parameter $\Xi$ \cite{boroudjerdi05a} for this system
is the same ($\Xi \approx 80$). For our model this is achieved for
$\lambda_B \approx 1.3$, where we do not yet observe aggregation. This
is due to the short length of our rods. For short rods counterion
condensation is much weaker \cite{antypov06a} due to end effects, and
therefore one needs stronger electrostatic interactions to observe
attraction and bundle formation.


\section{\label{conclusions}Conclusions:
}

We investigated the aggregation behavior of semi-flexible polyelectrolytes
with trivalent counterions. In dilute solutions the system shows a
micellization type aggregate size distribution. At low values of the Bjerrum
length single dispersed PEs are found, whereas at the other extreme of large
couplings macroscopic phase separation is observed. Our most important finding
is that finite size aggregates exist at thermodynamic equilibrium for a small
window of Bjerrum length values. The aggregate size distribution of these PE
bundles is rather broad. The finite size aggregates are a result of the
correlation of the trivalent counterions at short distances, and repulsive
electrostatic interactions at long distances. The entropic terms due to
counterions, and end-effects due to the finite length of the PE chains also
contribute to this delicate balance. The fraction of condensed counterions are
found to be significantly lower than those for an infinite PE rod, which leads
to an increased counterion entropy.

\section{Acknowledgements}
We would like to thank the EC for providing a Marie
Curie Intra-European Postdoctoral Fellowship to M. Sayar (MCIEF
500604). Additional funding from the DFG through grant SFB 625 and
Ho1108/11-3 is gratefully acknowledged. We thank M. Deserno and
K. Kremer for stimulating scientific discussions.

\end{document}